\documentclass[11pt]{article}
\usepackage{graphicx,color}
\usepackage{geometry}
\pdfoutput=1
\begin{document}

\small{
\begin{center}
 \textbf{Computational modelling of cancer evolution by multi-type branching processes}
\end{center}

\begin{center}
{Maroussia Slavtchova-Bojkova*}\\
{Sofia University "St. Kliment Ohridski" and Institute of Mathematics and Informatics at Bulgarian Academy of Sciences, Sofia, Bulgaria - bojkova@fmi.uni-sofia.bg}\\
\vspace{0.5cm}

{Kaloyan Vitanov}\\
{Sofia University "St. Kliment Ohridski",  Sofia, Bulgaria - kvitanov@uni-sofia.bg}\\

\end{center}

\begin{center}
{\bf Abstract}
\end{center}

\setlength{\parindent}{0pt}

Metastasis, the spread of cancer cells from a primary tumour to secondary location(s) in the human organism, is the ultimate cause of death for the majority of cancer patients. That is why, it is crucial to understand metastases evolution in order to successfully combat the disease. We consider a metastasized cancer cell population after medical treatment (e.g. chemotherapy). Arriving in a different environment the cancer cells may change their  lifespan and reproduction, thus they may proliferate into different types. If the treatment is effective, in the context of branching processes it means, the reproduction of cancer cells is such that the mean offspring of each cell is less than one (the branching process is called  subcritical). However, it is possible mutations to occur during cell division cycle. These mutations can produce a new cancer cell type, which is resistant to the treatment (having supercritical reproduction with mean offspring more than one cell). Cancer cells from this new type may lead to the rise of a non-extinction branching process. The above scenario leads us to the choice of a decomposable multi-type age-dependent branching process as a relevant framework for studying the asymptotic behavior of such complex structures. Our previous theoretical results  are related to the asymptotic behaviour of
 the waiting time until the first occurrence of a mutant starting a non-extinction process and the modified hazard function as a measure of immediate recurrence of cancer disease.
In the present paper these asymptotic results are used  for developing numerical schemes and algorithms implemented in Python via the NumPy package for approximate calculation of the corresponding quantities.  In conclusion, our conjecture is that this methodology can be advantageous in revealing the role of the lifespan distribution of the cancer cells in the context of cancer disease evolution and other complex cell population systems, in general. \\

{\bf Keywords}: Decomposable multi-type branching processes; Probability of extinction; Mutations; Waiting time to escape mutant; Modified hazard function.
}\\

\setlength{\parindent}{0pt}

{\bf 1. Introduction}

 In the last several decades branching processes proved to be appropriate models for describing cancer evolution as a particular case of  cell populations with complex structures.  
  For recent books, with emphasis on biological applications, see
Kimmel, M. \& Axelrod, D. (2002), Haccou et al. (2007) and also Durrett (2015), especially
for branching modeling in cancer. For a nice example of how branching processes can
be used to solve important problems in biology and medicine, the interesting  references are the papers of Iwasa et al. (2003), Iwasa et al. (2004). Classical references for branching processes theory are the books of Harris (1963), Athreya, K. \& Ney, P. (1972), Jagers (1975),
and Mode (1985).

 Two main consequences originating from the biological nature of cancer motivate us to consider a multi-type branching model in continuous time for studying proper measurable quantities connected to cancer. Firstly, the fact that there can be more than one type of metastasis in the human organism,  possibly after local elimination of the initial tumor by proper medical treatment. Secondly, arriving in a completely different environment the cancer  cell may change its characteristics concerning lifespan and division. Stemming from the multistage theory of cancer, our main idea in this study is to estimate the risk of cancer recurrence possibly after local elimination by means of multi-type decomposable branching  model, with $n$ types of cells, $n > 2$.
 This paper comes as a generalization of the results in the paper by Slavtchova-Bojkova et al. (2017), where  a two--type branching process in continuous time is used to model mutations occurring in a population of successfully treated cancer cells.

 In Section 2 we present our methodology based significantly on the theory of branching processes. We focus our attention on the so-called modified  hazard function, giving the quantification of the risk of immediate appearance of cancer, conditioned on the event it doesn't appeared by time $t$ and there are still cells of type $i = 0, \dots, n$. As in this case the quantities consisting in that function are satisfying non-linear integral equations, which are not of renewal type, our aim is to propose a method for solving such system of integral equations. In fact we propose a numerical method, which application and behaviour is illustrated  in Section 3 for  three specific setups of conditions imposed on the parameters driving the evolution of the model. Computational experiments are two-fold: to emphasize on the role of the lifespan distributions, from one point of view, and to support the theoretical considerations developed by Slavtchova-Bojkova, M. \& Vitanov, K. (2019), on the other side. More precisely, the choice of the offspring distributions and probabilities of mutations will be clarified in Section 3. We end up with discussion of the results and the contribution made by this research in Section 4.\\

{\bf 2. Methodology} \\

We will start with the definition of our multi-type Bellman-Harris branching process (MBHBP):
\begin{enumerate}
 \item There are $n+1$, $n > 1$, different types of cells, which are developing and reproducing independently form other existing cells in the population;

\item Each type $i$,  $i = 0, \dots, n$, has a (possibly) distinctive (continuous) distribution $G_i(t)  = P(\tau_i \le t),$ $G_i(0^{+}) = 0,$ of the lifespan $\tau_i$  and a (possibly) distinctive (discrete) distribution  $\{p_{ik}\}_{k=0}^{\infty}$, $  \sum_{k=0}^{\infty} p_{ik} = 1$, of the number of cells in the offspring $\nu_i.$
     We denote by $f_i(s) =   \sum_{k=0}^{\infty} p_{ik}s^k, |s| \le 1$  the probability generating function of the offspring $\nu_i$;

\item Each descendant of a   type-$i$  cell, $ i= 1,\dots, n$ can mutate at birth, independently of other cells, to any other type, with probabilities $u_{ik},$  $0 \le u_{ik} \le 1, k = 0, \dots, n$, $ \sum_{k=0}^n u_{ik} = 1$. Descendants of the   mutant type $0$ cannot mutate to another type, i. e. $u_{00}=1$, meaning also that there is no backward mutation, hence the model is decomposable;

\item Formally $ \Big \{ \mathbf{Z}(t) = \Big(Z^0(t), Z^1(t), \dots, Z^n(t) \Big) \Big\}_{t \ge 0},$ where $\{Z^i(t)\}_{t \ge 0}$ stands for the number of cells of type $i, i = 0,\dots, n$ at time $t$ respectively.
\end{enumerate}

\begin{figure}[htb!]
\centering
 \includegraphics[scale = 0.4]{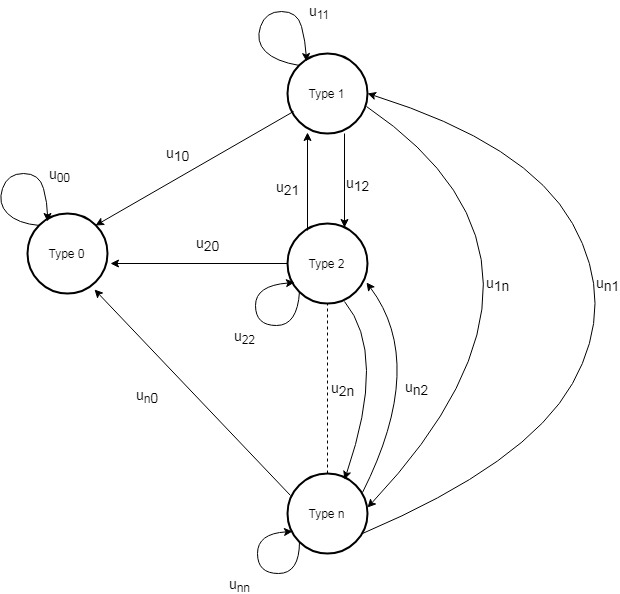}
  \caption{Flow diagram of transitions between types leading to the decomposable multi-type BHBP.}
  \label{plots_1}
  \end{figure}

A representation of the relationships and transitions between cells types
is given in Figure \ref{plots_1}.

{\bf Immediate risk of escaping extinction}

\indent One of the useful characteristics from application view point, associated with the occurrence of mutations in our MBHBP is the probability of occurrence of a ``successful'' mutant  within a very small interval $ dt $ after moment $ t,$ where by ``successful'' we mean a mutation leading to the indefinite survival of the cell population. More precisely,  this probability
 will be called ``immediate risk of escaping extinction''.

\indent  In our model, if there are no subcritical cells left in the cell population, the probability of occurrence of a ``successful'' mutant  is $0.$ Therefore we will investigate a modification of the standard formulation of the hazard function. Let us define a modified hazard function $ g_i (t) $ for each type $i = 1, \dots, n$ in the following way:
\begin{equation}\label{immediate risk}
g_i(t)dt = P\left( T_i {\in} (t,t+dt] | T_i>t, \sum_{j=1}^n Z^j(t)>0  \right ),
\end{equation}
where the random variables (r.v.) $T_i$ denote the waiting time of the occurrence of the first ``successful'' mutant, i.e., $T_i$ is the waiting time to the appearance of a process, which escapes extinction, provided that initially the process starts with one cell of type $i,$ $i= 1, \dots, n$.
In other words, we will consider the probability of occurrence of a ``successful'' mutant immediately after moment $ t $,   conditionally on the event that at moment $ t $ the population has at least one cell from an arbitrary $i,$ $i= 1, \dots, n$. \\
\indent From equation (\ref{immediate risk}) we derived
$$ 
g_i(t)dt = \frac{P( T_i {\in} (t,t+dt] | Z^i(0)=1, Z^m(0)=0, m{\neq} i )} {P(T_i>t , {\sum_{j=1}^n} Z^j(t)>0)  }
$$ 
$$
 = \frac{P( T_i {\in} (t,t+dt] | Z^i(0)=1, Z^m(0)=0, m \neq i)}{ Q_{i,t} - P(T_i>t, \sum_{j=1}^n Z^j(t)=0)},
$$

\noindent which shortly can be rewritten as:
\begin{equation}\label{m3}
g_i(t) = \frac{F_{T_i}^{(1)} (t)}{Q_{i,t} - V_{i,t}}.
\end{equation}

In the last equation (\ref{m3}) $F_{T_i}^{(1)}$ is the probability density function of $T_i$,  $Q_{i,t}$ is defined by

\begin{equation} \label{4}
Q_{i,t} \equiv P \left ( T_i > t \right ) = 1- G_i(t) + \displaystyle \int_0^t f_i (u_{i0}q_0 + \sum_{r=1}^n u_{ir}Q_{r,t-y}) dG_i(y), \ \ Q_{i,0} = 1,
\end{equation}
and $V_{i,t}$ satisfies
\begin{equation}\label{5}
V_{i,t} \equiv P\left ( T_i>t, \sum_{j=1}^n Z^j(t)=0 \right ) = \displaystyle \int_0^t f_i (u_{i0}q_0 + \sum_{r=1}^n u_{ir}V_{r,t-y}) dG_i(y), \ \ V_{i,0} = 0,
\end{equation}
as it is proven in Slavtchova-Bojkova, M., \& Vitanov, K. (2019), where  $q_0$ (the probability of extinction of type $0$ population) is the smallest non-negative root of the equation $f_0(s) = s.$

\indent From a practical point of view, we would like to  calculate $ g_i (t) $. Let us look at each of the components in the equation (\ref{m3}). The derivative  $F_{T_i}^{(1)} (t)$   in the nominator  can be approximated numerically using  forward difference approximation, i.e. for   partitioning the $ [0, T] $ interval with step $ h $,  we use points $ 0 = 0h, 1h, 2h, ..., Nh = T, (N + 1) h $, thus yielding
\begin{equation}\label{5m}
 F^{(1)}(kh)=\frac{F_T\big((k+1)h\big)-F_T(kh)}{h} + O(h),  k=0,1,...N.
\end{equation}

 {\bf Schemes for numerical calculations}

For the quantities in the denominator of (\ref{m3}) we have the following recurrent approximations:\\

\noindent I. Let $ t = 0 $. We have $V_{i,0}=0,$ $Q_{i,0} = 1,$ $i = 1, \dots, n$. \\

\noindent II. Let $ t = kh $. Note that for every $i = 1, \dots, n$ we can write

$$
\int_0^{kh} f_i\Big(u_{i0}s +  \sum_{r=1}^{n}u_{ir} Q_{r,(kh-y)}(s) \Big)dG_i(y)= $$

$$ = \sum_{j=1}^k  \int_{(j-1)h}^{jh}  f_i\Big(u_{i0}s + \sum_{r=1}^{n}u_{ir}Q_{r,(kh-y)}\Big)dG_i(y)
$$
 and approximating the integrals in the sum on the right-hand side in equation (\ref{4}), applying the right rectangle rule, we arrive at:
\begin{equation}\label{6}
Q_{i, kh} {\approx} 1-G_i(kh)+\sum_{j=1}^k f_i\Big(u_{i0}q_0+\sum_{r=1}^{n}u_{ir}Q_{r,(k-j)h}\Big) {\times} \Big(G_i(jh)-G_i\big((j-1)h \big) \Big).
\end{equation}

Analogously for $V_{i,kh},$ defined by (\ref{5}):
\begin{equation}\label{7}
V_{i,kh}{\approx} \sum_{j=1}^k f_i \Big(u_{i0}q_0 + \sum_{r=1}^{n}u_{ir}V_{r,(k-j)h}\Big){\times}\Big(G_i(jh)-G_i\big((j-1)h\big)\Big).
\end{equation}


{\bf 3. Results}

In this section we will present results obtained from calculations using equations (\ref{6}), (\ref{7}) and the scheme described by (\ref{5m}) done in three setups. These setups differ from each other in the type of distributions used for modeling the lifespan of the distinct cell types. In general, setup 1 (see Table 2) considers distributions which do not exhibit heavy--tails, setup 2 is the same as setup 1 except that a heavy--tailed distribution is used for one of the subcritical types, and setup 3 considers only heavy--tailed distributions. More precisely, in this paper, we will restrain ourselves with the following cases, where we model all cell types in setup 1 with exponential distributions, in setup 2 we will change the distribution of type 1 cells from exponential to lognormal, and in setup 3 we will model all cell types with lognormal distributions. In our full research we used numerous different combinations of exponential, truncated normal, gamma, lognormal, Pareto, Weibull and Cauchy distributions, throughout cell types $0,1,...,n$, all of those combinations yielded similar results as those stated below.
Also, our experiments concerned with step size $h$ have, so far, revealed that the results from calculations done with $h=0.1$ are practically identical to those obtained with $h=10^{-n}, n=2,3$.

For all calculations we use the same set of values for $f_i(s)$ and $u_{ij}$. We summarize the parameters for the three setups in the Table 1 below: \\

\begin{table}[ht]
\centering
\begin{tabular}{ | l l | c | c | c | c | c |}
\hline
 & $ f_i(s)$ & $f^{'}_i(1)$ & $u_{i0}$ & $u_{i1}$ & $u_{i2}$ & $u_{i3}$  \\ 
\hline
Type 0 & $0.2s^0 + 0.45s^2 + 0.35s^4$ & 2.3  &1&&& \\ 
\hline
Type 1 & $0.64s^0 + 0.36s^2$ &0.72& 0.05  &0.7 & 0.1 & 0.15  \\ 
\hline
Type 2 & $0.7s^0 + 0.12s^2+0.18s^4$ & 0.96& 0.1& 0.07 & 0.8 & 0.03 \\ 
\hline
Type 3 & $0.78s^0 + 0.22s^4$ & 0.88& 0.01& 0.07& 0.02 & 0.9 \\ 
\hline
\end{tabular}
\label{table setup 1}
\caption{Parameters of offspring distributions and mutation probabilities}
\end{table}

 In the next Table 2 we summarize the lifespan distributions for each setup and type:\\
\begin{table}[ht]
\centering
 \begin{tabular}{| l | c | c| c |}
\hline
 & $ setup 1$ & $setup 2$ & $setup 3$   \\ 
\hline
Type 0 & $Exp(2)$ & $Exp(2)$  & $Lognormal(0.5,1)$ \\ 
\hline
Type 1 & $Exp(3)$& $Lognormal(0.7,0.9)$ &$Lognormal(0.7,0.9)$ \\ 
\hline
Type 2 & $Exp(4.5)$ & $Exp(4.5)$& $Lognormal(0.35,1.1)$\\ 
\hline
Type 3 & $Exp(6)$& $Exp(6)$& $Lognormal(2.5,0.2)$ \\ 
\hline
\end{tabular}
\label{table setups}
\caption{Lifespan distributions used in numerical computations.}
\end{table}\\

{\bf Comparison of the modified hazard functions $g_i(t)$} \\

We proceed with  calculations regarding $g_i(t), i=1,2,3$. We let $h = 0.1$ for setups 1-3 and we begin with setup 1 with $t = 300$,
  setup 2 with $t = 500$ and setup 3 with  $t = 1000$.

\begin{figure}[htb!]
\centering
\minipage[t]{0.45\textwidth}
\centering
\includegraphics[height=0.25\textheight]{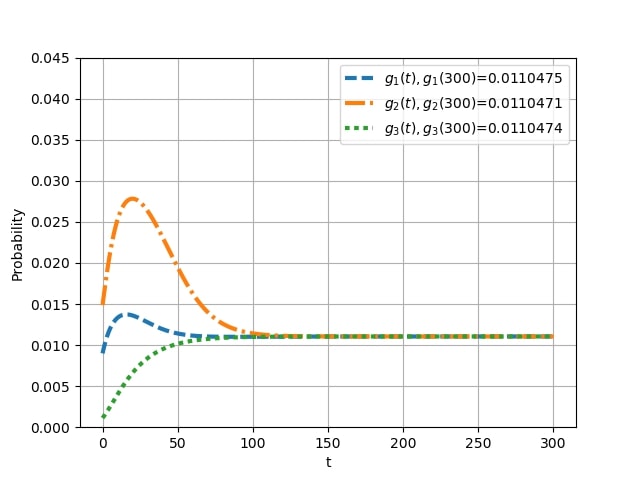}
\centering
\caption{Behaviour of $g_i(t)$ in setup 1.}
\label{83_setup_1}
\endminipage
\end{figure}

\begin{figure}[htb!]
\centering
\minipage[t]{0.45\textwidth}
\centering
\includegraphics[height=0.25\textheight]{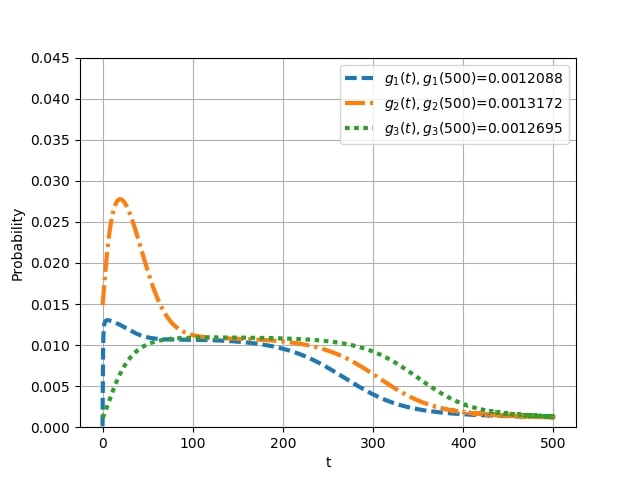}
\centering
\caption{Behaviour of $g_i(t)$ in setup 2.}
\label{83_setup_2}
\endminipage
\end{figure}

\begin{figure}[htb!]
\centering
\minipage[t]{0.45\textwidth}
\centering
\includegraphics[height=0.25\textheight]{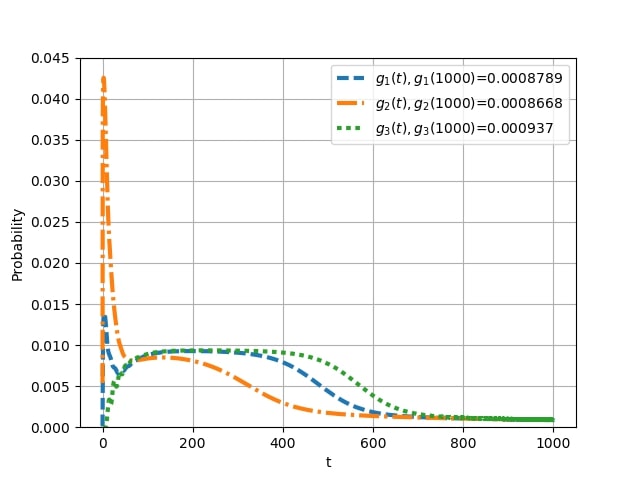} 
\endminipage\hfill
\centering
\caption{Behaviour of $g_i(t)$ in setup 3.} 
\label{83_setup_3}
\end{figure}

As it can be seen in Figure \ref{83_setup_2} and Figure \ref{83_setup_3}, introducing heavy-tailed distributions into the mix leads to a period of ``flatness'' in the values of $g_i(t), i = 1 ,2, 3$, followed by a period of monotone decreasing to  0. \\
\indent Let us now inspect the expression for $g_i(t)$. Considering the nominator, it is clear that it monotonically approaches 0 as $t \to \infty.$   On the other hand, for the denominator we have  ${P(T_i>t , {\sum_{j=1}^n} Z^j(t)>0 ) } \geq 0$, as it is a probability and obviously
  $P(T_i>t , {  \sum_{j=1}^n} Z^j(t)>0 )  > 0$.   \\

{\bf 4. Discussion and Conclusion}

Once again we state that the numerical results presented in the current paper are a subset of all  calculations we made using  exponential, truncated normal, gamma, lognormal, Pareto, Weibull and Cauchy distributions with various values for their parameters.  We admit, however, that until strict theoretical results are obtained, our numerical results (connected primarily with the lifespan distribution functions) may not be conclusive for all possible continuous distributions.\\
\indent As a conclusion we may summarize that:
\begin{enumerate}

\item We cannot guarantee that $g_i(t), i = 1, 2, 3$ converges to 0 for all heavy-tailed distributions (see  Figures \ref{83_setup_1} - \ref{83_setup_3}) at least without any additional conditions. It may be possible that $g_i(t)$ has non-zero limit value, as Figure \ref{83_setup_1}, Figure \ref{83_setup_2} and Figure \ref{83_setup_3}  suggest.

\item There are indications that introducing heavy-tailed distributions (recall Figure \ref{83_setup_2} and Figure \ref{83_setup_3} and note that the $Lognormal(0.7, 0.9)$ distribution has similar expected value and variance when compared to the $Exp(3)$ distribution) is influencing  on the behaviour of $g_i(t), i = 1, 2, 3,$ which is one of the conjectures we would like to check.

\end{enumerate}

{\bf Acknowledgements} The research was partially supported by the National Scientific Foundation of Bulgaria at the Ministry of Education and Science, grant No KP-6-H22/3, and Ministerio de Econom\'\i a y Competitividad and the FEDER
through the Plan Nacional de Investigaci\'on Cient\'\i fica, Desarrollo e Innovaci\'on Tecnol\'ogica,
grant MTM2015-70522-P, Spain. \\

{\bf References}\\

Athreya, K., \& Ney, P. (1972). Branching processes. Springer: New York.

Durrett, R. (2015). Branching Process Models of Cancer. Springer: Cham.

Haccou, P., Jagers, P.,  \& Vatutin, V. (2007). Branching processes: variation, growth
and extinction of populations. Cambridge, Cambridge University Press.

Harris, T. (1963). The theory of branching processes. Springer: Berlin.

Iwasa, Y., Michor, F., \&  Nowak, M. (2003). Evolutionary dynamics of escape from
biomedical intervention. Proc. Biol. Sci., B 270(1533), 2573--2578.

Iwasa, Y., Michor, F., \&  Nowak, M. (2004). Evolutionary dynamics of invasion and
escape. J. Theor. Biol., 226(2), 205--214.

Jagers, P. (1975). Branching Processes with Biological Applications. (1st ed.). John
Wiley \& Sons.

Kimmel, M.,  Axelrod, D. (2002). Branching Processes in Biology. Springer, New York.

Mode, C. (1985). Stochastic Processes In Demography and Their Computer Implementation. Springer: Berlin/Heidelberg.

Python Core Team (2015). PYTHON 3.5.2.: A dynamic, open source programming language. Python Software Foundation.

Slavtchova-Bojkova, M., Trayanov, P.,  \& Dimitrov, S. (2017). Branching processes in
continuous time as models of mutations: Computational approaches and algorithms. Comput. Stat.  Data Anal. 113,   111--124.
http://dx.doi.org/10.1016/j.csda.2016.12.013.

Slavtchova-Bojkova, M. \& Vitanov, K. (2019) Multi-type age-dependent branching processes as models of metastasis evolution, Stochastic Models, https://doi.org/10.1080/15326349.2019.1600410.

\end{document}